\documentstyle[11pt,epsf]{article}
\newcommand {\dfn} {\stackrel{\Delta} {=}}
\newcommand {\exe} {\stackrel{\cdot} {=}}

\newcommand {\reals} {{\rm I\!R}}

\newcommand {\bx} {\mbox{\boldmath $x$}}
\newcommand {\by} {\mbox{\boldmath $y$}}
\newcommand {\bz} {\mbox{\boldmath $z$}}

\newcommand {\bE} {\mbox{\boldmath $E$}}

\newcommand {\blambda} {\mbox{\small \boldmath $\lambda$}}

\newcommand {\bX} {\mbox{\boldmath $X$}}
\newcommand {\bY} {\mbox{\boldmath $Y$}}

\newcommand{\calA}{{\cal A}}

\newcommand{\calC}{{\cal C}}

\newcommand{\calN}{{\cal N}}

\newcommand{\calX}{{\cal X}}
\newcommand{\calY}{{\cal Y}}
\newcommand{\calZ}{{\cal Z}}
\begin{document}
\thispagestyle{empty}
\title{Optimum Estimation via Gradients of Partition Functions
and Information Measures: A Statistical--Mechanical Perspective}
\author{Neri Merhav}
\date{}
\maketitle

\begin{center}
Department of Electrical Engineering \\
Technion -- Israel Institute of Technology \\
Technion City, Haifa 32000, Israel\\
{\tt merhav@ee.technion.ac.il}\\
\end{center}
\vspace{1.5\baselineskip}
\setlength{\baselineskip}{1.5\baselineskip}

\begin{abstract}
In continuation to a recent work on the 
statistical--mechanical analysis of minimum mean square error
(MMSE) estimation in Gaussian noise via its relation 
to the mutual information (the I--MMSE relation), here we 
propose a simple and more direct relationship between optimum estimation
and certain information measures (e.g., the information density
and the Fisher information), which can be viewed as
partition functions and hence are amenable to analysis using statistical--mechanical
techniques. The proposed approach has several advantages, most notably,
its applicability to general sources and channels, as opposed to the
I--MMSE relation and its variants which hold only for certain classes of channels
(e.g., additive white Gaussian noise channels). We then demonstrate the derivation
of the conditional mean estimator and the MMSE in a few examples. Two of these
examples turn out to be generalizable to a fairly wide class of sources and
channels. For this class, the proposed approach is shown to yield an 
approximate conditional mean estimator and an MMSE
formula that has the flavor of a single--letter expression. 
We also show how our approach can easily be generalized to situations of mismatched
estimation.\\

\noindent
{\bf Index Terms:} Conditional mean estimation, minimum mean squared error,
partition function, statistical mechanics, Fisher information.
\end{abstract}

\section{Introduction}

Relationships between signal estimation, signal detection, and information measures,
both in discrete time and continuous time, have been
known for decades \cite{Bucy79},\cite{Duncan70},\cite{Kailath70}
and have gained a remarkable degree of revived interest and
research activity in the
last several years, see, e.g., 
\cite{Guo09}, \cite{GSV05a}, \cite{GSV05b}, \cite{GSV08},
\cite{PV06}, \cite{RC09}, \cite{Verdu09} and references therein.

In particular, in \cite{GSV05a}, Guo, Shamai and Verd\'u have derived a
relation between the mutual information between the input and the output
of an additive white Gaussian noise (AWGN) channel and
the minimum mean squared error (MMSE) of non--causal estimation of the channel input
based on its output. In particular, this relation, which is often called
the I--MMSE relation, shows that the derivative of the mutual information
with respect to (w.r.t.) the signal--to--noise (SNR) is equal to half of the
MMSE, and it
is intimately related to the de Bruijn
identity \cite[Sec.\ 17.7]{CT06}. Later, this relation has been generalized
and further developed
in several directions: Guo, Shamai, and Verd\'u \cite{GSV05b} 
and Raginsky and Coleman \cite{RC09} have derived relations of the same spirit 
for more general additive channels. 
Palomar and Verd\'u \cite{PV06} have studied relations between the covariance
matrix of the MMSE estimator 
and arbitrary gradients of the mutual information for a general vector Gaussian
channel, which allows also a linear transformation of the input signal.
In \cite{GSV08}, relations
between information measures and estimation measures have been derived for Poisson channels. 
More recently, Verd\'u \cite{Verdu09} extended the I--MMSE relation of Gaussian noise 
to the paradigm of mismatched conditional mean estimation, that is,
to deal with an estimator that is optimally matched to a wrong probability
distribution assumed on the input signal. The excess mean squared error (MSE) due to
this mismatch was shown to be related to the Kullback--Leibler divergence
between the channel output distributions corresponding 
to the true and the assumed input distributions (see also \cite{Guo09} for a
further study in this direction). In \cite{MGS10}, the I--MMSE relation was
further investigated from a statistical physics perspective, where among other
results, it was
demonstrated how statistical--mechanical tools can be harnessed in order to
assess the MMSE via the I--MMSE relation of \cite{GSV05a}, using the fact
that in many cases, the mutual information can be viewed as the partition function of
a certain physical system.

This paper is a further development in the above described direction of \cite{MGS10}. 
The main idea is that, for the purpose of evaluating the covariance matrix of the MMSE
estimator, one may use a conceptually simple and more direct relationship
between the MMSE covariance matrix and other information measures, that can also be
presented in the form of a certain partition function and hence be analyzed using methods of
statistical physics. The main advantage of the proposed approach, over those of
the I--MMSE relations and its variants, is its full generality: It applies, in
principle, to any joint probability function $P(\bx,\by)$ of the channel input signal
$\bx=(x_1,\ldots,x_n)$, to be estimated, and the channel output
$\by=(y_1,\ldots,y_m)$ (where $m$ and $n$ are positive integers), 
provided that certain technical regularity conditions
hold. The channel $P(\by|\bx)$ does not even have to be additive, as opposed
to the assumptions made in
\cite{GSV05b} and \cite{RC09}. Moreover, the dimension $m$ of the channel
output vector $\by$ does not have to be the same as the dimension $n$ of the input
vector $\bx$.

In a nutshell, the idea is to define, for a given $n$--vector of real--valued
parameters $\blambda=(\lambda_1,\ldots,\lambda_n)$, the `partition function'
$$Z(\by,\blambda)=\sum_{\bx}
\exp\left\{\sum_{i=1}^n\lambda_ix_i\right\}P(\bx,\by),$$
where we have implicitly assumed that $\bx$ takes on discrete values,
otherwise, the sum should simply be replaced by an integral. Now, it is
straightforward to show that the gradient of $\ln Z(\by,\blambda)$ w.r.t.\
$\blambda$, computed at $\blambda=0$, gives the conditional mean estimator
$\hat{\bx}=\bE\{\bX|\by\}$, whereas the expectation of the Hessian of the same
function, again, at $\blambda=0$, gives the error covariance matrix of the
MMSE estimator. As we shall see in the sequel, $\ln Z(\by,\blambda)$ lends
itself to closed form analytic evaluation (in the spirit of a single--letter
formula) in a fairly wide spectrum of situations,
using methods of statistical mechanics. Thus, the MMSE estimator and its
performance can quite easily be derived too in these situations. Moreover,
as was demonstrated extensively in \cite{MGS10}, the statistical--mechanical
perspective on estimation--theoretic problems, may offer, not only analysis
techniques, but also some important insights with regard to threshold
effects (whenever existent) via the inspection of possible {\it phase transitions} in the
parallel statistical--mechanical model.

Besides the general applicability of this approach, it has
several additional advantages:
\begin{enumerate}
\item As mentioned in the previous
paragraph, it provides, not only the MMSE error covariance matrix, but also the
conditional mean estimator itself.
\item As will be seen, several variants of these relations between estimation
measures and information measures can be offered. In some cases, one of the
relations may be more convenient to work with than the others.
\item The approach is easy to extend to the mismatched case. Furthermore, it 
allows mismatch in both the source and the channel (as opposed to
\cite{Verdu09}, which allows mismatch in the source only).
\end{enumerate}

The remaining part of this paper is organized as follows. In Section 2, we
establish notation conventions. In Section 3, we first derive the basic relations between
the conditional mean estimator, as well as its error covariance matrix,
and the above--mentioned partition function.
In the same section, we also discuss this relation
and derive a few variants that involve also
information measures, like the information density, the Fisher
information, etc. We also outline the extension to mismatched estimation.
In Section 4, we provide three examples. In Section 5,
we show how two of them set the stage to the analysis of a more
general class of joint distributions, $P(\bx,\by)$. 
Finally, in Section 6, we summarize and conclude the paper.

\section{Notation Conventions}

Throughout this paper, scalar random
variables (RV's) will be denoted by capital
letters, their sample values will be denoted by
the respective lower case letters, and their alphabets will be denoted
by the respective calligraphic letters.
A similar convention will apply to
random vectors and their sample values,
which will be denoted with same symbols in the bold face font.
Thus, for example, $\bX$ 
will denote a random vector $(X_1,\ldots,X_n)$,
and $\bx=(x_1,\ldots,x_n)$ is a specific vector value in $\calX^n$,
the $n$-th Cartesian power of $\calX$. The
notations $y_i^j$ and $Y_i^j$, where $i$
and $j$ are integers and $i\le j$, will designate segments $(y_i,\ldots,y_j)$
and $(Y_i,\ldots,Y_j)$, respectively. 

Probability functions 
will be denoted generically by the letter $P$ or $Q$.
In particular, $P(\bx,\by)$ is the joint probability mass function  
(in the discrete case) or the joint density (in the continuous case)
of the desired channel input vector $\bx=(x_1,\ldots,x_n)$ and the observed channel
output vector $\by=(y_1,\ldots,y_m)$. Accordingly, $P(\bx)$ will denote the
marginal of $\bx$, $P(\by|\bx)$ will denote the conditional probability mass
(or density) of $\by$ given $\bx$, induced by the channel, and so on.
Whenever there is room for ambiguity, these probability functions will be
subscripted by the names of the random variables and the conditionings,
according to standard notation conventions in probability theory and
information theory. Throughout the sequel, we will assume discrete valued
alphabets, mostly for the sake of simplicity and convenience. Extensions to continuous valued
situations will be straightforward with summations being replaced by integrations,
etc. Indeed, some of our examples will involve continuous valued random
variables. 

The expectation operator of a generic function $f(\bx,\by)$ w.r.t.\ the joint
distribution $P$ of
$(\bX,\bY)$ will be denoted by $\bE\{f(\bX,\bY)\}$. The conditional expectation
of the same function given that $\bY=\by$, denoted
$\bE\{f(\bX,\bY)|\bY=\by\}$, and which is obviously identical to
$\bE\{f(\bX,\by)|\bY=\by\}$, is, of course, a function of $\by$. 
On substituting $\bY$ in this function, this becomes then a random variable
which will be denoted by
$\bE\{f(\bX,\bY)|\bY\}$. When using vectors and matrices in a
linear--algebraic format, $n$--dimensional vectors, like $\bx$ (and $\bX$),
will be understood as column vectors, the operator $(\cdot)^T$ will denote
vector or matrix transposition, and so, $\bx^T$ would be a row vector.
For two positive sequences $\{a_n\}$ and $\{b_n\}$, the notation
$a_n\exe b_n$ means equivalence in the exponential order, i.e.,
$\lim_{n\to\infty}\frac{1}{n}\log(a_n/b_n)=0$. Finally, the indicator function of an
event $\calA$ will be denoted by $1\{\calA\}$. I.e., $1\{\calA\}=1$ is $\calA$
occurs, and $1\{\calZ\}=0$ if not.

\section{MMSE Estimation Relations}

This section consists of two subsections. In the first,
we derive the main basic relations and in the second,
we show how to extend the scope to the case of mismatched estimation.

\subsection{Basic Relations}

Let $\bX=(X_1,\ldots,X_n)$,
and $\bY=(Y_1,\ldots,Y_m)$ ($n$ and $m$ being
positive integers),
be two random vectors,
jointly distributed according to a given probability function $P(\bx,\by)$. 
It is further assumed that the alphabet $\calX$, of each component of $\bX$, consists of
a set of real valued numbers, i.e., $\calX\subseteq \reals$. This assumption
is obviously necessary in order to make the problem of estimating $\bX$, in the MSE sense,
a meaningful problem. The conditional mean estimator of $\bX$ based on $\bY$,
i.e., $\hat{\bX}=\bE\{\bX|\bY\}$ is well--known to be the optimum estimator in
the MSE sense, i.e., it minimizes the MSE $\bE\{(X_i-\hat{X}_i)^2\}$ for all
$i=1,2,\ldots,n$. The MMSE in estimating $X_i$ is then
$\bE\{(X_i-\bE\{X_i|\bY\})^2\}$, i.e., the expected conditional variance of
$X_i$ given $\bY$. More generally, the MMSE error covariance matrix $E$ is
an $n\times n$ matrix whose $(i,j)$--th element is given by
$\bE\{(X_i-\bE\{X_i|\bY\})(X_j-\bE\{X_j|\bY\})\}$. This matrix can be represented as
the expectation (w.r.t.\ $\bY$) of the conditional covariance matrix of $\bX$ given
$\bY$, henceforth denoted $\mbox{Cov}\{\bX|\bY\}$. I.e.,
$$E=\bE\{\mbox{Cov}\{\bX|\bY\}\}=\bE\{\bX\bX^T\}-\bE\{\bE\{\bX|\bY\}\cdot\bE\{\bX^T|\bY\}\}.$$

Defining a column vector of $n$ real valued 
parameters, $\blambda=(\lambda_1,\ldots,\lambda_n)^T$,
consider the following
function:
$$Z(\by,\blambda)\dfn
\sum_{\bx\in\calX^n}
\exp\{\blambda^T\bx\}P(\bx,\by)=
\sum_{\bx\in\calX^n}
\exp\{\blambda^T\bx\}P(\bx)P(\by|\bx),$$
where it is assumed that the sum (or integral, in the
continuous case) converges uniformly at least in some
neighborhood of $\blambda=0$.\footnote{If this assumption is not met, one can
instead, parametrize each component $\lambda_i$ of $\blambda$ as a purely
imaginary number $\lambda_i=j\omega_i$ ($j=\sqrt{-1}$), as is done in the definition of the
characteristic function.}
It is straightforward to see now that:
\begin{equation}
\label{1st}
\frac{\partial\ln
Z(\by|\blambda)}{\partial\lambda_i}\bigg|_{\blambda=0}=
\frac{\sum_{\bx\in\calX^n}x_i P(\bx,\by)}{P(\by)}=\sum_{x_i\in\calX}x_iP(x_i|\by)
=\bE\{X_i|\by\},
\end{equation}
i.e., 
\begin{equation}
\label{grad}
\bE\{\bX|\by\}=
\nabla_{\blambda}\ln Z(\by,\blambda),
\end{equation}
where $\nabla_{\blambda}$ denotes the gradient w.r.t.\ $\blambda$.
Similarly, upon taking second order derivatives, one obtains
$$\frac{\partial^2\ln
Z(\by|\blambda)}{\partial\lambda_i\partial\lambda_i}\bigg|_{\blambda=0}=
E\{X_iX_j|\by\}-\bE\{X_i|\by\}\cdot\bE\{X_i|\by\}=\mbox{Cov}\{X_i,X_j|\by\},$$
and so,
\begin{equation}
\label{hess}
E=\bE\left\{\nabla_{\blambda}^2\ln
Z(\bY,\blambda)\bigg|_{\blambda=0}\right\},
\end{equation}
where $\nabla_\lambda^2$ is the Hessian w.r.t.\ $\blambda$, namely, the matrix
of second order derivatives w.r.t.\ pairs of components of $\blambda$.
Note that here and throughout the sequel, we will always refer to gradients and Hessians
of functions w.r.t.\ $\blambda$, computed at the point $\blambda=0$. It will
therefore be convenient to use,
for a generic function $g$, the shorthand notations
$\nabla_0 g(\blambda)$ and $\nabla_0^2g(\blambda)$ to designate
$\nabla_{\blambda}g(\blambda)\bigg|_{\blambda=0}$ and
$\nabla_{\blambda}^2g(\blambda)\bigg|_{\blambda=0}$, respectively.

Another, perhaps simpler, way to look at the relations (\ref{grad}) and
(\ref{hess}) is the
following: Obviously, for a given $\by$, $M(\by,\blambda)=\sum_{\bx}
e^{\blambda^T\bx} P(\bx|\by)$ is the moment generating function pertaining
to the conditional distribution of $\bx$ given $\by$ and so, its derivatives
relative to $\{\lambda_i\}$, computed at $\blambda=0$, yield the conditional moments
$\bE\{X_i|\by\}$,
$\bE\{X_i^2|\by\}$,
$\bE\{X_iX_j|\by\}$, etc. Therefore, $\ln M(\by,\blambda)$ is a
generator of the corresponding conditional cumulants,
$\bE\{X_i|\by\}$, $\mbox{Var}\{X_i|\by\}$,
$\mbox{Cov}\{X_i,X_j|\by\}$, etc. Now, observe that $\ln M(\by,\blambda)$ differs from
$\ln Z(\by,\blambda)$ merely by the additive term $\ln P(\by)$, which does
not depend on $\blambda$ anyway and hence does not affect the gradient and
Hessian w.r.t.\
$\blambda$. Therefore, $\ln Z(\by,\blambda)$ is a generator of conditional
cumulants, exactly like $\ln M(\by,\blambda)$. 
An important point, however, is that we prefer $\ln Z(\by,\blambda)$ over
$\ln M(\by,\blambda)$ because normally, it is more convenient to work with 
the joint distribution $P(\bx,\by)$ (or equivalently, with
the source $P(\bx)$ and forward channel $P(\by|\bx)$) rather than with
the backward channel (or the posterior) $P(\bx|\by)$.\footnote{
As a side remark, we shall mention also
the physical perspective: if $Z(\by,\blambda)$ is thought of as the
partition function of a certain statistical--mechanical model (as discussed in
the Introduction), where the components of $\blambda$ are thought of as
certain generalized forces or fields that are acting on the individual particles, then the above
relation between the second order derivative of 
$\ln Z(\by,\blambda)$ w.r.t.\ $\lambda_i$ and $\lambda_j$ and the
(conditional) covariances between the corresponding state variables,
$X_i$ and $X_j$, is known as one of the versions of the
fluctuation--dissipation theorem in statistical mechanics \cite[p.\ 32, eq.\
(2.44)]{MM09}, which
relates between the linear response of the system (to an infinitesimally small
perturbation in its parameters) and its fluctuations in equilibrium.}

We next derive several alternative versions of this relation between the
error covariance matrix of the MMSE estimator and derivatives of $\ln Z$.
First, observe that $Z(\by,\blambda)$ is proportional to
$P_{\blambda}(\by)\cdot\Theta(\blambda)$, where
$$\Theta(\blambda)=\sum_{\bx\in\calX^n}P(\bx)\exp\{\blambda^T\bx\}$$
and $P_{\blambda}(\by)$ is the output marginal of $\by$ induced by the channel 
$P(\by|\bx)$ and the modified source distribution
$P_{\blambda}(\bx)\dfn e^{\blambda^T\bx}P(\bx)/\Theta(\blambda)$.
We therefore obtain
\begin{eqnarray}
E &=&
\bE\left\{\nabla_0^2\ln
Z(\bY,\blambda)\right\}\nonumber\\
&=&\bE\left\{\nabla_0^2\ln[P_{\blambda}(\bY)\cdot
\Theta(\blambda)]\right\}\nonumber\\
&=&\nabla_0^2\ln \Theta(\blambda)+
\bE\left\{\nabla_0^2\ln P_{\blambda}(\bY)
\right\}\nonumber\\
&=&\mbox{Cov}\{\bX\}-J,
\end{eqnarray}
where
$\mbox{Cov}\{\bX\}=\bE\{\bX\bX^T\}-\bE\{\bX\}\cdot\bE\{\bX^T\}$
is the covariance matrix of $\bX$ and $J$ is the Fisher information matrix of
estimating $\blambda$ based on $\bY$, computed at the point $\blambda=0$. The Fisher
information matrix $J$ can also be expressed as
$$J=\bE\left\{\nabla_0\ln P_{\blambda}(\bY)\cdot\nabla_0^T\ln
P(\bY|\blambda)\right\}.$$
Equivalently, we obtained
$$J=\mbox{Cov}\{\bX\}-E=\bE\{\bE\{\bX|\bY\}\cdot\bE\{
\bX^T|\bY\}\}.$$
Note that $J$ can also be obtained as the negative expectation of the Hessian (or,
equivalently, as
the covariance matrix of the gradient) of the {\it information density}
\cite{VH94},
$$i_{\blambda}(\bx;\by)=\ln[P(\by|\bx)/P_{\blambda}(\by)],$$ 
which is again, computed at
$\blambda=0$.

Sometimes it is more convenient to square the first derivative of $\ln Z$
than to take the second derivative. In these cases, the following relationship
may be useful:
\begin{eqnarray}
\Xi&\dfn&\bE\left\{[\nabla_0\ln Z(\bY,\blambda)]\cdot
[\nabla_0\ln Z(\bY,\blambda)]^T\right\}\nonumber\\
&=&\bE\left\{[\nabla_0\ln \{P(\bY|\blambda)\cdot\Theta(\blambda)\}]\cdot
[\nabla_0\ln
\{P(\bY|\blambda)\cdot\Theta(\blambda)\}]^T
\right\}\nonumber\\
&=&\bE\left\{[\nabla_0\ln P(\bY|\blambda)]\cdot
[\nabla_0\ln P(\bY|\blambda)]^T\right\}+
[\nabla_0\ln\Theta(\blambda)]\cdot
[\nabla_0\ln \Theta(\blambda)]^T\nonumber\\
&=&J+\bE\{\bX\}\cdot\bE\{\bX^T\}\nonumber\\
&=&\mbox{Cov}\{\bX\}+\bE\{\bX\}\cdot\bE\{\bX^T\}-E\nonumber\\
&=&\bE\{\bX\bX^T\}-E
\end{eqnarray}
and so,
$$E=\bE\{\bX\bX^T\}-\Xi.$$
Particularizing these results to the MMSE, 
$$\mbox{mmse}(\bX|\bY)\dfn\sum_{i=1}^n\bE\{(X_i-\bE\{X_i|\bY\})^2\},$$
which is the trace of $E$, we
have the following relations, which we formulate as a proposition.\\

\noindent
{\bf Proposition 1.} The following formulas for the MMSE hold:
\begin{eqnarray}
\mbox{mmse}(\bX|\bY)&=&\sum_{i=1}^n\bE\left\{\frac{\partial^2\ln
Z(\bY,\blambda)}{\partial\lambda_i^2}\bigg|_{\blambda=0}\right\}\\
&=&\sum_{i=1}^n\left[\mbox{Var}\{X_i\}+\bE\left\{\frac{\partial^2\ln
P(\bY|\blambda)}{\partial\lambda_i^2}\bigg|_{\blambda=0}\right\}\right]\\
&=&\sum_{i=1}^n\left[\mbox{Var}\{X_i\}-\bE\left\{\left[\frac{\partial\ln
P(\bY|\blambda)}{\partial\lambda_i}\right]^2\bigg|_{\blambda=0}\right\}\right]\\
&=&\sum_{i=1}^n\left[\bE\{X_i^2\}-\bE\left\{\left[\frac{\partial\ln
Z(\bY,\blambda)}{\partial\lambda_i}\right]^2\bigg|_{\blambda=0}\right\}\right]
\end{eqnarray}
In the second and the third formulas,
$\ln P(\bY|\blambda)$ can be replaced by $\ln i(\bX;\bY)$, 
thus relating the MMSE to the information density.\\

\subsection{Extension to the Mismatched Case}

In this short subsection, we are outlining how our approach can easily
be extended to handle situations of mismatched estimation.
Consider a mismatched estimator
which is the conditional mean of $\bX$ given $\bY$, based on an
incorrect joint distribution $Q(\bx,\by)$,
whereas the true joint distribution continues to be $P(\bx,\by)$.
Denoting by $Z_P(\by,\blambda)$ and $Z_Q(\by,\blambda)$
the corresponding partition
functions, and by $\bE_P$ and $\bE_Q$, the
corresponding expectations,
our approach can easily be generalized to handle this case
as follows:
\begin{eqnarray}
E&=&\bE_P\left\{(\bX-\bE_Q\{\bX|\bY\})(\bX^T-\bE_Q\{\bX^T|\bY\})\right\}\nonumber\\
&=&\bE_P\{\bX\bX^T\}-
\bE_P\{\bE_P\{\bX|\bY\}\bE_Q\{\bX^T|\bY\}\}-\nonumber\\
& &\bE_P\{\bE_Q\{\bX|\bY\}\bE_P\{\bX^T|\bY\}\}+
\bE_P\{\bE_Q\{\bX|\bY\}\bE_Q\{\bX^T|\bY\}\}\nonumber\\
&=&\bE_P\{\bX\bX^T\}-\bE_P\{[\nabla_0\ln
Z_P(\bY,\blambda)]\cdot[\nabla_0\ln Z_Q(\bY,\blambda)]^T\}-\nonumber\\
& &\bE_P\{[\nabla_0\ln
Z_Q(\bY,\blambda)]\cdot[\nabla_0\ln Z_P(\bY,\blambda)]^T\}
+\bE_P\{[\nabla_0\ln
Z_Q(\bY,\blambda)]\cdot[\nabla_0\ln Z_Q(\bY,\blambda)]^T\}.\nonumber
\end{eqnarray}
Thus, in particular, the MSE associated with the mismatched estimator
is given by
\begin{eqnarray}
\mbox{mse}_Q(\bX|\bY)&=&\sum_{i=1}^n\left[\bE_P\{X_i^2\}-
2\bE_P\left\{\frac{\partial\ln
Z_P(\bY,\blambda)}{\partial\lambda_i}\bigg|_{\blambda=0}\cdot
\frac{\partial\ln
Z_Q(\bY,\blambda)}{\partial\lambda_i}\bigg|_{\blambda=0}\right\}\right.\nonumber\\
& &+\left.\bE_P\left\{\left[\frac{\partial\ln
Z_Q(\bY,\blambda)}{\partial\lambda_i}\bigg|_{\blambda=0}\right]^2\right\}\right].
\end{eqnarray}

\section{Examples}

In this section, we provide three examples, where we show how the
log--partition function, $\ln Z(\by,\blambda)$, can be evaluated
for large $n$, using methods of statistical mechanics. Using the
relations derived in Subsection 3.1, we then show how the conditional
mean estimator and the MMSE can be approximated for large $n$.

\subsection{Example 1 -- A Codeword Transmitted Over an AWGN}

Our first example is taken from \cite[Subsection 5.2]{MGS10}, but here
we demonstrate how to derive the conditional mean estimator and the MMSE
using Proposition 1, rather than the I--MMSE relation. 
For the sake of completeness and convenience, we provide
here the full necessary details
(with the appropriate
modifications to accommodate the
method proposed herein), including those that already appear in \cite{MGS10}. As noted 
in \cite{MGS10}, the analysis of this model is intimately related to one of the
statistical mechanical techniques used in the analysis of the so called random energy
model (REM) of disordered magnetic materials, a.k.a.\ spin glasses in the
statistical physics literature (see
references in \cite{MGS10}).

Let $\bX$ be chosen uniformly at random from a codebook 
$\calC=\{\bx_0,\bx_1,\ldots,\bx_{M-1}\}$ of size $M=e^{nR}$. 
The codebook itself is also selected at random (and then revealed to the
estimator) in the following manner:
Each $\bx_i$ is
selected independently and uniformly at random from the surface of a sphere of radius
$\sqrt{nP_x}$ centered at the origin. The channel $P(\by|\bx)$ is an AWGN
channel (hence $m=n$) whose
noise variance is $1/\beta$ (keeping the same notation as in \cite{MGS10}). I.e.,
$$P(\by|\bx)=\left(\frac{\beta}{2\pi}\right)^{n/2}\exp\left\{\frac{\beta}{2}\|\by-\bx\|^2\right\}.$$
Thus, for a given $\by$, we have:
\begin{eqnarray}
Z(\by,\lambda)&=&\sum_{\bx\in\calC}
e^{-nR}\exp\{-\beta\|\by-\bx\|^2/2+\blambda^T\bx\}\nonumber\\
&=&e^{-nR}\exp[-\beta\|\by-\bx_0\|^2/2+\blambda^T\bx_0]+
\sum_{\bx\in\calC\setminus\{\bx_0\}}e^{-nR}
\exp[-\beta\|\by-\bx\|^2/2+\blambda^T\bx]\nonumber\\
&\dfn& Z_c(\by,\blambda)+Z_e(\by,\blambda),
\end{eqnarray}
where, without loss of generality, $\bx_0$ designates the transmitted codeword.
Now, since $\|\by-\bx_0\|^2$ is typically around $n/\beta$, $Z_c(\by,\blambda)$
would typically be about $e^{-nR}e^{-\beta\cdot n/(2\beta)}
e^{\blambda^T\bx_0}=e^{-n(R+1/2)+\blambda^T\bx_0}$.
As for $Z_e(\by,\blambda)$, we have:
$$Z_e(\by,\blambda)\exe e^{-nR}\int_{\reals}\mbox{d}\epsilon 
N(\epsilon)e^{-\beta
n\epsilon},$$
where $N(\epsilon)$ is the number of codewords $\{\bx\}$ in $\calC-\{\bx_0\}$
for which
$\|\by-\bx\|^2/2-\blambda^T\bx/\beta\approx n\epsilon$, namely, between $n\epsilon$ and
$n(\epsilon+\mbox{d}\epsilon)$.
Now, given $\by$,
$N(\epsilon)=\sum_{i=1}^M1\{\bx_i:~\|\by-\bx_i\|^2/2-\blambda^T\bx/\beta\approx
n\epsilon\}$
is the sum of $M$ i.i.d.\ Bernoulli
random variables and so, its expectation is
\begin{equation}
\overline{N(\epsilon)}=\sum_{i=1}^M\mbox{Pr}\{
\|\by-\bX_i\|^2/2-\blambda^T\bX_i/\beta\approx n\epsilon\}=
e^{nR}\mbox{Pr}\{\|\by-\bX_1\|^2/2-\blambda^T\bX_1/\beta\approx n\epsilon\}.
\end{equation}
Denoting
$P_y=\frac{1}{n}\sum_{i=1}^ny_i^2$ (typically, $P_y$ is about $P_x+1/\beta$),
the event $\|\by-\bx\|^2/2-\blambda^T\bx/\beta\approx n\epsilon$ is equivalent to the event
$\bx^T(\by+\blambda/\beta)\approx[(P_x+P_y)/2-\epsilon]n$
or equivalently,
$$\rho(\bx,\by)\dfn
\frac{\bx^T(\by+\lambda/\beta)}{n\sqrt{P_xP_y'}}\approx\frac{\frac{1}{2}(P_x+P_y)-\epsilon}
{\sqrt{P_xP_y'}}\dfn\frac{P_a-\epsilon}{P_g'},$$
where have defined $P_a=(P_x+P_y)/2$ and $P_g'=\sqrt{P_xP_y'}$,
where $P_y'=\frac{1}{n}\sum_i(y_i+\lambda_i/\beta)^2$.
The probability that a randomly chosen vector $\bX$ on the sphere would have
an empirical correlation coefficient $\rho$ with a given vector
$\by'=\by+\blambda/\beta$
(that is, $\bX$ falls within a cone of half angle $\arccos(\rho)$ around
$\by'$)
is exponentially $\exp[\frac{n}{2}\ln(1-\rho^2)]$.
For convenience, let us define
\[
\Gamma(\rho) = \frac12 \ln \left( 1- \rho^2 \right)
\]
so that we can write
$$\mbox{Pr}\{\|\by-\bX_1\|^2/2-\blambda^T\bX_1/\beta\approx n\epsilon\}\exe
\exp\left\{n \, \Gamma\left(\frac{P_a-
\epsilon}{P_g'}\right)\right\}.$$
If $\epsilon$ is such that
$$\Gamma\left(\frac{P_a-\epsilon}{P_g'}\right) > - R,$$
then the energy level $\epsilon$ will be typically populated with an
exponential
number of codewords, concentrated very strongly around its mean
$$\overline{N(\epsilon)}\exe\exp\left\{n\left[R+
\Gamma\left(\frac{P_a-\epsilon}{P_g'}\right)\right]\right\},$$
otherwise (which means that $\overline{N(\epsilon)}$ is exponentially small),
the energy level $\epsilon$ will not be populated by any codewords typically.
This
means that the populated energy levels range between
$$\epsilon_1\dfn P_a-P_g'\sqrt{1-e^{-2R}}$$
and
$$\epsilon_2\dfn P_a+P_g'\sqrt{1-e^{-2R}},$$
or equivalently, the populated values of $\rho$ range between $-\rho_*$ and
$+\rho_*$
where $\rho_*=\sqrt{1-e^{-2R}}$. By large deviations and saddle--point
methods,
it follows that for a typical realization of the randomly chosen code, we have
\begin{eqnarray}
Z_e(\by,\blambda)
&\exe&e^{-nR}\max_{\epsilon\in[\epsilon_1,\epsilon_2]}
\exp\left\{n\left[R+
\Gamma\left(\frac{P_a-\epsilon}{P_g'}\right)-\beta\epsilon
\right]\right\}\nonumber\\
&=&\max_{\epsilon\in[\epsilon_1,\epsilon_2]}
\exp\left\{n\left[\Gamma\left(\frac{P_a-\epsilon}{P_g'}\right)-\beta\epsilon
\right]\right\}\nonumber\\
&=&
\exp\left\{n\left[\max_{|\rho|\le\rho_*}\left\{\frac12\ln(1-\rho^2)-\beta(P_a-\rho
P_g')
\right\}\right]\right\}\ .
\end{eqnarray}
The derivative of $\frac{1}{2}\ln(1-\rho^2)+\rho\beta P_g'$ w.r.t.\ $\rho$
vanishes within $[-1,1]$ at:
$$\rho=\rho_\beta\dfn\sqrt{1+\theta^2}-\theta$$
where
$$\theta\dfn\frac{1}{2\beta P_g'}.$$
This is the maximizer as long as $\sqrt{1+\theta^2}-\theta\le\rho_*$, namely,
$\theta >e^{-2R}/{2\rho_*}$, or equivalently, $\beta<\rho_* e^{2R}/P_g'$, which
for
$P_g'=\sqrt{P_x(P_x+1/\beta)}$ ($\|\blambda\|$ is small), is equivalent to
$\beta<\beta_R\dfn
(e^{2R}-1)/P_x$.
Thus, for the typical code we have
$$Z_e(\beta|\by)\exe
\left\{\begin{array}{ll}
\exp\left\{n\left[\frac{1}{2}
\ln(1-\rho_\beta^2)-\beta (P_a -\rho_\beta P_g')\right]\right\}, & \beta < \beta_R\\
\exp\{-n[R+\beta(P_a-\rho_* P_g')]\}, & \beta \ge \beta_R .
\end{array}\right.$$
Taking now into account $Z_c(\by,\blambda)$, it is easy to see that for
$\beta\ge\beta_R$
(which means $R<C$), $Z_c(\by,\blambda)$ dominates $Z_e(\by,\blambda)$, whereas for
$\beta<\beta_R$ it is the other way around. It follows then that
$$Z(\by,\blambda)\exe
\left\{\begin{array}{ll}
\exp\left\{n\left[\frac{1}{2}\ln(1-\rho_\beta^2)-\beta (P_a -\rho_\beta
P_g')\right]\right\}, 
& \beta < \beta_R\\
\exp\left\{-n\left(R+\frac{1}{2}\right)+
\blambda^T\bx_0\right\}, & \beta \ge \beta_R .
\end{array}\right.$$
A very similar analysis applies also to the derivative
$\frac{\partial}{\partial\lambda_i}\ln Z(\by,\blambda)$, which is essentially a weighted
average of $x_i$ with weights proportional to 
$\overline{N(\epsilon)}e^{-\beta\epsilon}$ for all
$\epsilon\in[\epsilon_1,\epsilon_2]$. Thus, the exponentially dominant weight is due to the
term that maximizes the exponent. Assuming that the correct codeword $\bx_0$
is dominant ($Z_c >> Z_e$, which is the case when $R<C$), 
this weighted average is obviously dominated by
the $i$--th component of $\bx_0$, in which case the MMSE essentially vanishes. 
Otherwise, for $R > C$, $Z_e$ dominates the partition function and the
weighted average is overwhelmingly dominated by the term corresponding to the
maximizing $\epsilon$, or equivalently, the maximizing $\rho$, which is
$\rho_\beta$. This means that the conditional mean estimator of $X_i$ is
approximately given by:
\begin{eqnarray}
\bE\{X_i|\by\}&\approx&
\frac{\partial}{\partial\lambda_i} 
\left[\frac{n}{2}\ln(1-\rho_\beta^2)+\beta\rho_\beta
nP_g'\right]\nonumber\\
&=&-\frac{n\rho_\beta}{1-\rho_\beta^2}
\frac{\partial\rho_\beta}{\partial\lambda_i}+
\beta nP_g'\frac{\partial\rho_\beta}{\partial\lambda_i}
+\beta\rho_\beta n\frac{\partial P_g'}{\partial\lambda_i}
\nonumber\\
&=&n\frac{\partial\rho_\beta}
{\partial\lambda_i}\left(-\frac{\rho_\beta}{1-\rho_\beta^2}
+\beta P_g'\right)
+\beta n\rho_\beta\frac{\partial P_g'}{\partial\lambda_i}
\nonumber\\
&=&
n\beta\rho_\beta\frac{\partial P_g'}{\partial\lambda_i}
\nonumber\\
&=&\beta\rho_\beta n\cdot\frac{\sqrt{P_x}}{2\sqrt{P_y}}\cdot
\frac{2y_i}{\beta n}\nonumber\\
&=&\rho_\beta\sqrt{\frac{P_x}{P_x+1/\beta}}
\cdot y_i\nonumber\\
&=&\frac{P_x}{P_x+1/\beta}\cdot y_i,
\end{eqnarray}
where in the last step we have used the identity
$\rho_\beta=\sqrt{P_x/(P_x+1/\beta)}$, which can easily be verified.
This is simply the linear Wiener estimator that would have been applied
had the input been zero--mean, i.i.d.\ Gaussian, with variance $1/\beta$
(see also \cite{MGS10}). 
According to Proposition 1, the MMSE associated with $X_i$ is given by
$$\bE\{(X_i-\bE\{X_i|\bY\})^2\}\approx P_x-\bE\{\bE^2(X_i|\bY)\}=
P_x-\left(\frac{P_x}{P_x+1/\beta}\right)^2\cdot(P_x+1/\beta)=\frac{P_x}{1+\beta
P_x},$$
as expected.

\subsection{Example 2 -- The Curie--Weiss Model}

Consider a binary source
$$P(\bx)=C_n\exp\left\{\frac{a}{2n}\left(\sum_{i=1}^nx_i\right)^2+b\sum_{i=1}^nx_i\right\}
~~~~\bx\in\{-1,+1\}^n$$
where $a$ and $b$ are parameters and $C_n$ is a normalization constant, which
is immaterial for our purposes (as it is going to disappear upon taking
derivatives w.r.t.\ $\{\lambda_i\}$, and the same comment applies to the
constants $C_n'$ and $C_n''$ below).
Let the channel be binary and symmetric, i.e.,
$$P(y|x)=\frac{e^{\beta xy}}{2\cosh(\beta)},~~~~y\in\{-1,+1\}.$$
Then, the partition function $Z(\by,\blambda)$ can be represented as a
one--dimensional integral using the Hubbard--Stratonovich transform,
which in turn can be assessed using saddle point methods, as is frequently
done in the statistical physics literature. Specifically, we have the
following:
\begin{eqnarray}
Z(\by,\blambda)&=&C_n'\sum_{\bx}\exp\left\{\frac{a}{2n}\left(\sum_{i=1}^nx_i\right)^2+
b\sum_{i=1}^nx_i+\beta\sum_{i=1}^n
x_iy_i+\sum_{i=1}^n\lambda_ix_i\right\}\\
&=&C_n'\sum_{\bx}\exp\left\{\sum_{i=1}^nx_i(\beta y_i+\lambda_i+b)+
\frac{a}{2n}\left(\sum_{i=1}^nx_i\right)^2
\right\}\\
&=&C_n''\sum_{\bx}\exp\left\{\sum_{i=1}^nx_i(\beta
y_i+\lambda_i+b)\right\}\cdot\int_{-\infty}^{+\infty}\mbox{d}\theta
\exp\left\{-\frac{n\theta^2}{2a}+\theta\sum_{i=1}^nx_i
\right\}\\
&=&C_n''\int_{-\infty}^{+\infty}\mbox{d}\theta e^{-n\theta^2/(2a)}\sum_{\bx}
\exp\left\{\sum_{i=1}^nx_i(\beta
y_i+\lambda_i+b+\theta)\right\}\\
&=&C_n''\int_{-\infty}^{+\infty}\mbox{d}\theta e^{-n\theta^2/(2a)}
\prod_{i=1}^n[2\cosh(\beta
y_i+\lambda_i+b+\theta)]\\
&=&2^nC_n''\int_{-\infty}^{+\infty}\mbox{d}\theta
\exp\left\{-\frac{n\theta^2}{2a}+
\sum_{i=1}^n\ln\cosh(\beta
y_i+\lambda_i+b+\theta)\right\}.
\end{eqnarray}
Thus,
\begin{eqnarray}
\frac{\partial\ln Z(\by,\blambda)}{\partial\lambda_i}&=&
\frac{\int_{-\infty}^{+\infty}\mbox{d}\theta\tanh(\beta
y_i+\lambda_i+b+\theta)\exp\left\{-\frac{n\theta^2}{2a}+
\sum_{i=1}^n\ln\cosh(\beta
y_i+\lambda_i+b+\theta)\right\}}{
\int_{-\infty}^{+\infty}\mbox{d}\theta \exp\left\{-\frac{n\theta^2}{2a}+
\sum_{i=1}^n\ln\cosh(\beta
y_i+\lambda_i+b+\theta)\right\}}\nonumber\\
&\approx&\tanh(\beta y_i+\lambda_i+b+\theta_*),
\end{eqnarray}
where $\theta_*$ is the maximizer of the expression at the exponent, i.e., it
is the solution to the zero--derivative equation:
$$\theta=\frac{a}{n}\sum_{i=1}^n\tanh(\beta y_i+\lambda_i+b+\theta).$$
Thus, the MMSE estimator is:
\begin{eqnarray}
\bE\{X_i|\by\}&=&
\frac{\int_{-\infty}^{+\infty}\mbox{d}\theta\tanh(\beta
y_i+b+\theta)\exp\left\{-\frac{n\theta^2}{2a}+
\sum_{i=1}^n\ln\cosh(\beta
y_i+b+\theta)\right\}}{
\int_{-\infty}^{+\infty}\mbox{d}\theta \exp\left\{-\frac{n\theta^2}{2a}+
\sum_{i=1}^n\ln\cosh(\beta
y_i+b+\theta)\right\}}\\
&\approx&\tanh(\beta y_i+b+\theta_*),
\end{eqnarray}
where now $\theta_*$ is understood to be taken with $\blambda=0$.
For $b\ne 0$, the asymptotic MMSE is then given by
$$\lim_{n\to\infty}\frac{\mbox{mmse}(\bX|\bY)}{n}=1-\bE\{\tanh^2(\beta
Y+b+\theta_0)\},$$
where $\theta_0$ is the solution to the equation
$$\theta=a\bE\{\tanh(\beta Y+b+\theta)\},$$
and where $Y$ is a binary $\{\pm 1\}$ RV, with mean $m^*\tanh(\beta)$,
$m^*$ being the dominant solution
to the equation $m=\tanh(am+b)$, i.e., the maximizer of
$h_2((1+m)/2)+am^2/2+bm$, where $h_2(\cdot)$ is the binary entropy function.
When $b=0$, $\theta_0$ becomes a
random variable which takes on, with equal probabilities, one of two values, each one
being the solution to the above displayed equation, except that in one of
them $Y$ has mean $m^*\tanh(\beta)$ and in the other, its mean is
$-m^*\tanh(\beta)$. 

This calculation is intimately related to the Curie--Weiss
model of magnetic spins \cite[Subsection 2.5.2, pp.\ 40--44]{MM09}, where the parameter $m$ plays the
role of magnetization.

\subsection{Example 3 -- The Generalized Multivariate Cauchy Noise Model}

Let $X_i\sim\calN(0,\sigma^2)$ be i.i.d.\ RV's, and let
the additive noise have a generalized multivariate Cauchy distribution,
i.e.,
$$P(\by|\bx)=\frac{C_{n,k}}{[1+(\by-\bx)^TS(\by-\bx)]^k}$$
where $C_{n,k}$ is a normalization constant,
$S$ is a positive definite matrix, and $k > 0$ is chosen large
enough (as a function of $n$) such $\int_{\reals^n}
\mbox{d}\bz/[1+\bz^TS\bz]^k < \infty$, i.e.,
$k > n/2$. The choice $k=(n+1)/2$ corresponds to
the ordinary multivariate Cauchy distribution. Here, however, we will
require moreover 
that $k$ is even large enough such that the second moments exist, 
i.e., $\int_{\reals^n}
\mbox{d}\bz\cdot \bz^T\bz/[1+\bz^TS\bz]^k < \infty$, which means $k > n/2+1$.
For simplicity, we will take $S$ to be the identity matrix. However, our
analysis easily extends to a general positive matrix $S$, as well as to a
general Gaussian vector $\bX$, not necessarily with i.i.d.\ components.
Using the Laplace transform identity $\int_0^\infty \mbox{d}t\cdot 
t^{k-1}e^{-st}
=\Gamma(k)/s^k$, we have:
\begin{eqnarray}
Z(\by,\blambda)&=&\int_{\reals^n}\mbox{d}\bx P(\bx)e^{\blambda^T\bx}
\cdot\frac{C_{n,k}}{[1+\sum_{i=1}^n(y_i-x_i)^2]^k}\\
&=&C_{n,k}\int_{\reals^n}\mbox{d}\bx P(\bx)e^{\blambda^T\bx}
\int_0^\infty\mbox{d}t\cdot\frac{t^{k-1}}{\Gamma(k)}\cdot
e^{-t[1+\sum_i(y_i-x_i)^2]}\\
&=&C_{n,k}'\int_0^\infty\mbox{d}t\cdot t^{k-1}e^{-t}
\int_{\reals^n}\mbox{d}\bx P(\bx)e^{\blambda^T\bx}\cdot
e^{-t\sum_i(y_i-x_i)^2}\\
&=&C_{n,k}''\int_0^\infty\mbox{d}t\cdot t^{k-1}e^{-t}
\prod_{i=1}^n\int_{\reals}\mbox{d}x_i e^{-x_i^2/2\sigma^2}e^{\lambda_ix_i}\cdot
e^{-t(y_i-x_i)^2}\\
&=&C_{n,k}''\int_0^\infty\mbox{d}t\cdot t^{k-1}e^{-t}
\left(\frac{2\pi\sigma^2}{1+2t\sigma^2}\right)^{n/2}\cdot
\exp\left\{-t\sum_{i=1}^ny_i^2+\sum_{i=1}^n
\frac{(ty_i+\lambda_i/2)^2}{t+1/2\sigma^2}\right\}.
\end{eqnarray}
and so,
$$\frac{\partial\ln Z(\by,\blambda)}{\partial\lambda_i}\bigg|_{\blambda=0}=
\frac{\int_0^\infty\mbox{d}t\cdot\frac{ty_i}{t+1/2\sigma^2}e^{-t}t^{k-1}
\exp\left\{-\frac{n}{2}\ln(1+2t\sigma^2)-
\frac{t}{1+2t\sigma^2}\sum_iy_i^2\right\}}{
\int_0^\infty\mbox{d}te^{-t}t^{k-1}
\exp\left\{-\frac{n}{2}\ln(1+2t\sigma^2)-
\frac{t}{1+2t\sigma^2}\sum_iy_i^2\right\}}$$
which can be approximated by $\hat{t}y_i/(\hat{t}+1/2\sigma^2)$,
where $\hat{t}$ is the value of $t$ that dominates the integral, i.e.,
$$\hat{t}=\mbox{argmax}_t\left[(k-1)\ln t-\frac{n}{2}\ln(1+2t\sigma^2)-
\frac{t}{1+2t\sigma^2}\sum_iy_i^2\right].$$
The derivation of the MMSE can be done in a similar manner as in the previous
examples.\\

\section{Joint Distributions with Generalized Spherical Symmetry}

Examples 2 and 3 of the previous section have one idea in common.
In both of them we expressed either the source or the channel as
a one--dimensional integral over a variable ($t$ or $\theta$, in those
examples), where for each value of this variable, we have a product form
measure, which enables, after applying 
saddle point analysis on this integral, to pass to a closed--form
formula, which has the flavor of a single--letter characterization.
In this section, we generalize this idea to establish a somewhat more general
framework.

Suppose that $m=n$ and the joint distribution of $\bX$ and $\bY$ is of the
form
$$P(\bx,\by)=F_n(\sum_i\phi(x_i,y_i)).$$ 
Let $f_n(t)$ be the inverse
Laplace transform of $F_n(s)$. Then, we have
\begin{eqnarray}
Z(\by,\blambda)&=&\int_{\reals^n}\mbox{d}\bx
e^{\blambda^T\bx}P(\bx,\by)\nonumber\\
&=&\int_{\reals^n}\mbox{d}\bx
e^{\blambda^T\bx}\int_0^\infty\mbox{d}tf_n(t)
\exp\left\{-t\sum_i\phi(x_i,y_i)\right\}\nonumber\\
&=&\int_0^\infty\mbox{d}t
f_n(t)\int_{\reals^n}\mbox{d}\bx
e^{\blambda^T\bx}\exp\left\{-t
\sum_i\phi(x_i,y_i)\right\}\nonumber\\
&=&\int_0^\infty\mbox{d}t
f_n(t)\prod_i\int_{\reals}\mbox{d}x_i
e^{\lambda_ix_i}\exp\{-t\phi(x_i,y_i)\}.
\end{eqnarray}
Before proceeding, we should
note that by using the Laplace transform, we have essentially represented
the joint distribution of $\bX$ and $\bY$ as a mixture of product form
measures, indexed by $t$, each being 
proportional to $\exp\{-t\sum_i\phi(x_i,y_i)\}$. If we normalize
these measures by $Z_t^n=[\sum_{x\in\calX}\sum_{y\in\calY}\exp\{-t\phi(x,y)\}]^n$, and define
the i.i.d.\ probability distribution
$$P(\bx,\by|t)=\frac{\exp\{-t\sum_i\phi(x_i,y_i)\}}{Z_t^n}$$
then $P(\bx,\by)$ is essentially expressed here as a mixture of i.i.d.\ probability
functions $\{P(\bx,\by|t)\}$, where $t$ can be thought of as a random
parameter whose prior is given by $w_n(t)=f_n(t)Z_t^n$. However, it
should be kept in mind that this integral representation goes somewhat further
than being a mixture of i.i.d.\ distributions because $f_n(t)$, and hence also
$w_n(t)$, may be negative for some ranges of $t$ even
when $F_n(s)$ is strictly positive for all $s$. For example, recall that
the inverse Laplace transform of $s^2/(s^2+\alpha^2)$ is $\sin(\alpha t)$
($t\ge 0$), and so, for
$F_n(s)=\alpha^n/(s^2+\alpha^2)^n$, $f_n(t)$ is given by the $n$--fold
convolution of $\sin(\alpha t)$ with itself. In such cases, $P(\bx,\by)$
cannot be considered a mixture of i.i.d.\ distributions.

Let us now denote
$$\rho(\lambda,y,t)=\ln\left[\int_{-\infty}^{\infty}\mbox{d}x e^{\lambda
x-t\phi(x,y)}\right],$$
$$\rho_0(y,t)=\rho(0,y,t)=\ln\left[\int_{-\infty}^{\infty}\mbox{d}x
e^{-t\phi(x,y)}\right],$$
and
$$\zeta(y,t)=\frac{\partial\rho(\lambda,y,t)}{\partial\lambda}\bigg|_{\lambda=0}
=\frac{\int_{\reals}\mbox{d}x\cdot x e^{-t\phi(x,y)}}
{\int_{\reals}\mbox{d}x\cdot e^{-t\phi(x,y)}}.$$
Then,
$$Z(\by,\blambda)=\int_0^\infty\mbox{d}t
f_n(t)e^{\sum_i\rho(\lambda_i,y_i,t)},$$
and so,
$$\bE\{X_i|\by\}=\frac{\partial\ln
Z(\by,\blambda)}{\partial\lambda_i}\bigg|_{\blambda=0}=
\frac{\int_0^\infty\mbox{d}t
f_n(t)\zeta(y_i,t)e^{\sum_i\rho_0(y_i,t)}}
{\int_0^\infty\mbox{d}t
f_n(t)e^{\sum_i\rho_0(y_i,t)}}$$
which is approximated by $\zeta(y_i,\hat{t})$, where
$\hat{t}$ is the maximizer of the expression
$$\ln|f_n(t)|+\sum_i\rho_0(y_i,t).$$
The MMSE of estimating $X_i$ is given by 
$$\mbox{mmse}(X_i|\bY)\approx\bE\{X_i^2\}-
\bE\{\zeta^2(Y_i,t_0(t))\}$$
where the second term is computed as follows:
$$\bE\{\zeta^2(Y_i,t_0(t))\}
=\int_0^\infty\mbox{d}t
w_n(t)
\bE\{\zeta^2(Y_i,t_0(t))|t\}$$
with the inner expectation being
$$\bE\{\zeta^2(Y_i,t_0(t))|t\}
=\frac{\int_{\reals}\mbox{d}ye^{\rho_0(y,t)}
\zeta^2(y,t_0(t))}
{\int_{\reals}\mbox{d}ye^{\rho_0(y,t)}}$$
and with $t_0(t)$ being the
value of $t'$ that maximizes
$$\left[\ln|f_n(t')|+n\cdot
\frac{\int_{\reals}\mbox{d}y\cdot e^{\rho_0(y,t)}\rho(y,t')}
{\int_{\reals}\mbox{d}y\cdot e^{\rho_0(y,t)}}
\right].$$
Thus, we have characterized both the conditional mean estimator and the MMSE
in the spirit of a single--letter formula for this class of joint
distributions.

The following further extensions of this formalism are conceptually
straightforward:
\begin{enumerate}
\item The range of the variable $t$ may not necessarily be $[0,\infty)$.
Our above analysis applies to whatever range as long as the integrals exist.
\item The joint distribution $P(\bx,\by)$ may be a function of more than one
statistic $\sum_i\phi(x_i,y_i)$, i.e.,
$$P(\bx,\by)=F_n\left(\sum_{i=1}^n\phi_1(x_i,y_i),\ldots,\sum_{i=1}^n\phi_k(x_i,y_i)\right).$$
In this case, one may apply a Laplace transform of a higher dimension
$$F(s_1,\ldots,s_k)=\int_0^\infty\cdot\cdot\cdot\int_0^\infty\mbox{d}t_1\cdot\cdot\cdot\mbox{d}t_k
f(t_1,\ldots,t_k)e^{-s_1t_1-\ldots-s_kt_k},$$
 where $s_i=\sum_i\phi_i(x_i)$,
$i=1,2,\ldots,k$.
\item The assumption that the $i$--th term of $\sum_i\phi(x_i,y_i)$ depends
only on the $i$--th coordinate of $\by$ is not really necessary. The
derivation continues to hold, for example, if we allow more generally
the form $\sum_i\phi(x_i,y_i,y_{i-1},\ldots,y_{i-k})$.
\item The case where  $\phi$ is a quadratic form can be extended to allow
a quadratic form that involves all coordinates of $\bx$ and $\by$
collectively,
using a positive definite matrix $S$ for weighting. In other words, joint distributions with
elliptic symmetry are allowed, with the form $P(\bx,\by)=F_n[(\bx,\by)^TS(\bx,\by)]$, where
$(\bx,\by)$ denotes the concatenated column vector
of dimension $(n+m)$ formed by $\bx$ and $\by$, and the matrix $S$
is of dimension $(n+m)\times (n+m)$. In this case, the kernel is Gaussian
and hence the estimator is linear for a given $t$. 
\end{enumerate}

\section{Conclusion}

In this paper, we have proposed a simple relation between MMSE estimation
measures and a certain expression, which can be viewed as a partition
function, and hence be analyzed using methods of statistical mechanics.
This partition function is also related to several information measures,
like the information density and the Fisher information. The proposed approach
has several advantages over the I--MMSE relation and its variants:
\begin{enumerate}
\item It is conceptually simple and direct.
\item It applies in full generality, for every joint distribution of the
desired random vector $\bX$ and its noisy observation vector $\bY$.
\item It provides, not only the MMSE error covariance matrix, but also 
the conditional mean estimator itself $\hat{\bx}=\bE\{\bX|\by\}$.
\item It offers several alternative expressions of the MMSE (see Proposition 1).
\item The approach is easy to extend to the mismatched case and it allows
mismatch, not only in the marginal of $\bX$, but in the entire joint density
$P(\bx,\by)$.
\end{enumerate}
Finally, considering earlier work on the I--MMSE
relation and its various variants that were discussed in the Introduction,
it would be natural to seek relations between MMSE estimation to the Hessian
of the mutual information. One can show, using the same
techniques as in Subsection 3.1, that the following relation holds:
$$E=\nabla_0^2 I_{\blambda}(\bX;\bY)+\mbox{Cov}\{\bX\}-
\mbox{Cov}\left\{(\bX-\bE\{\bX\})(\bX-\bE\{\bX\})^T,\ln\frac{P(\bY|\bX)}{Z(\bY,\blambda)}\right\},$$
where $I_{\lambda}(\bX;\bY)$ is the mutual information induced by the joint
distribution
$$P_{\blambda}(\bx,\by)=\frac{e^{\blambda^T\bx}P(\bx,\by)}{\Theta(\blambda)}.$$
Unfortunately, this relation seems somewhat more complicated 
and not as useful as the I--MMSE relation of \cite{GSV05a} or the relations proposed in
Subsection 3.1 herein.



\end{document}